\documentclass[preprintnumbers,prd,twocolumn,showpacs,floatfix,preprintnumbers,letterpaper,amsmath,amssymb,nofootinbib,superscriptaddress]{revtex4}
\usepackage{graphicx}
\usepackage{epsfig}
\usepackage{bm}
\usepackage{amsfonts}
\begin{document}
 
\title{Can structure formation distinguish $\Lambda CDM$ from Non-minimal $f(R)$ Gravity?}

\author{ Shruti Thakur }  
\email {shruti.thkr@gmail.com}
\affiliation{ Department of Physics and Astrophysics, University of Delhi, Delhi-110007, India}

\author{ Anjan A Sen }
\email {aasen@jmi.ac.in}
\affiliation{ Center for Theoretical Physics, Jamia Millia Islamia, New Delhi-110025, India}

\date{\today}

\begin{abstract}
Non-minimally coupled $f(R)$ gravity model is an interesting approach to explain the late time acceleration of the Universe without introducing any exotic matter component in the energy budget of the Universe. But distinguishing such model with the concordance $\Lambda$CDM model using present observational data is a  serious challenge. In this paper, we address this issue using the observations related to the growth of matter over density $g(z)$ as measured by different galaxy surveys. As background cosmology is not sufficient to distinguish  different dark energy models, we first find out the functional form for $f(R)$ ( which is non-minimally coupled to the matter lagrangian) that produces the similar background cosmology as in $\Lambda$CDM.  Subsequently we calculate the growth for the matter over density $g(z)$ for such non minimally coupled $f(R)$ models and compare them with the $\Lambda$CDM Universe. We also use the measurements of $g(z)\sigma_{8}(z)$  by different galaxy surveys to reconstruct the behavior for $g(z)$ and $\sigma_{8}(z)$ for both the non-minimally coupled $f(R)$ gravity models as well as for the $\Lambda$CDM. Our results show that there is a small but finite window where one can distinguish the non-minimally coupled $f(R)$ models with the concordance $\Lambda$CDM.
\end{abstract}

\maketitle
\section{Introduction}
The quest to explore gravity theory beyond Einstein's general theory relativity (GTR) on large cosmological scales, in recent years, has been aroused by the prospect of explaining the late time acceleration of the Universe. It has now been established beyond any doubt that our Universe is currently going through an accelerated expanding phase which has been started in recent past \cite{obs}. The theoretical approach to explain such an acceleration can be broadly divided into two categories. In one approach, one has to add some exotic component with negative pressure (known as dark energy) in the energy budget of the Universe as normal matter or radiation component can not initiate accelerated expansion.  Simplest of such component, known as cosmological constant ( with equation of state $w=-1$) can explain all the available cosmological data but at the same time is plagued by the embarrassing problem of fine tuning  and cosmic coincidence \cite{fintun}. One can also consider more exotic scalar fields \cite{scalar0,scalar1,scalar2,scalar3,scalar4,scalar5,scalar6,scalar7,scalar8,scalar9,scalar10,
scalar11,scalar12,scalar13,scalar14,scalar15,scalar16,scalar17} to mimic the required negative pressure which may solve at least the cosmic coincidence problem and can have far more striking cosmological consequences. Unfortunately the presently available cosmological data can not distinguish these two models decisively.

The other approach is to give up the idea of adding extra component in energy budget of the Universe but to modify the gravity theory on large cosmological scales. The idea is to look for departure from the GTR that can effectively mimic a cosmological constant and result the necessary acceleration. $f(R)$ gravity theory \cite{fr} is one such example where instead of linear dependence of the Einstein-Hilbert (EH) action on the Ricci scalar, one includes nonlinear dependence. It is particularly interesting and has been studied extensively. Recently a generalization of such theory has been proposed  which involves a non-minimal coupling between curvature and matter \cite{nonmin}. In this model, one extends the presence of non-linear functions of the scalar curvature in  the EH action by incorporating an additional  term involving the coupling between the Ricci scalar and the matter. Astrophysical and cosmological signatures of such non-minimally coupled $f(R)$ gravity models have been studied by various authors \cite{nonmin} (see also \cite{tomi} for similar investigation) .

However describing background cosmological evolution for any particular model is not sufficient to remove the observed degeneracies between different modified gravity models and $\Lambda$CDM. Growth of matter perturbation is extremely important in this regard as the evolution equations for growth of matter perturbation are completely different in two scenarios. This can result characteristic signatures in CMB as well as in matter power spectrum from galaxy clusters which in turn can help to remove the degeneracies between the two models. In a recent paper \cite{berto} the evolution of cosmological perturbations in the presence of non-minimal coupling between curvature and matter has been studied.

In this work, we extend the work done in \cite{berto} to the observational front. We study how far the present cluster data can distinguish the model with non-minimal coupling between curvature and matter from the standard $\Lambda$CDM model. For this we proceed in a two stage manner. We first look for the possible form for the $f(R)$ that can give rise to the same background evolution as in $\Lambda$CDM Universe. After reconstructing the suitable form for the $f(R)$, we calculate the matter power spectrum for such a model and see whether the currently available observational data can distinguish this model from the $\Lambda$CDM.

\section{The Non-Minimal Coupling in Modified Gravity Theories }

\vspace{2mm}
The action for non minimally coupled $f(R)$ gravity models is given by
\begin{equation}
S= \int \left( \frac{R}{2 \kappa^2}+(1+f(R))L_m \right) \sqrt{-g}d^4x,
\label{action}
\end{equation}

\noindent
where $\kappa^2 = 8\pi G$, $f(R)$ is a dimensionless function of the scalar curvature $R$ and ${\cal L}_{m}$ is the matter lagrangian.

Due to the presence of coupling between the curvature and the matter, there will be an energy transfer between them which can be seen in the corresponding conservation equation given by
\begin{equation}
\nabla^{\mu}T_{\mu\nu} = \frac{df/dR}{1+f(R)}(g_{\mu\nu}L_m-T_{\mu\nu})\nabla^{\mu}R.
\end{equation}

The non vanishing term in the r.h.s of the above equation shows the energy transfer  between curvature and matter. This exchange of energy between the curvature and matter is a special feature in non minimally coupled $f(R)$ gravity models. But in a homogenous and isotropic Universe, the energy momentum tensor for the matter should have a perfect fluid form and once we assume the form of the matter lagrangian as ${\cal L}_{m} = -\rho_{m}$, \cite{lagrange}($\rho_{m}$ being the energy density for the matter), it is easy to show that $\nabla^{\mu}T_{\mu\nu} = 0$ is satisfied for the background evolution.

Varying the action (\ref{action}) with respect to the metric tensor $g_{\mu\nu}$, one obtains the Einstein's equation as

\begin{equation}
\chi G_{\mu\nu}= \frac{1}{2}R(1-\chi)g_{\mu\nu} +\chi_{,\mu;\nu}-g_{\mu\nu} \Box \chi+\kappa^2 (1+f(R))T_{\mu\nu}  \label{eins}
\end{equation}
where $\chi=1+2\kappa^2L_m \frac{df}{dR} $.

\noindent
Assuming a spatially flat Friedmann-Robertson-Walker spacetime with scale factor a(t)
\begin{equation}
ds^2 = -dt^2+a(t)^2 \delta_{ij}dx^idx^j
\end{equation}
 the time component of above equation is 
\begin{equation}
 H\dot{\chi} = (\dot{H}+2H^2)\chi -\frac{R}{6}-H^2 \chi +\frac{\kappa^2}{3}\rho_m(1+f(R)).
 \label{zero}
 \end{equation}

Our first goal is to look for those models that mimic the background evolution for the $\Lambda$CDM Universe. We know the form of the Hubble parameter for the $\Lambda$CDM Universe which is given by
\begin{equation}
h^2 = \frac{H^2}{H_0^2}=\frac{\Omega_{m0}}{a^3}+\Omega_{\Lambda 0},
\label{hub0}
\end{equation}

where $H_{0}$,  $\Omega_{m0}$ and $\Omega_{\Lambda 0}$  are the Hubble parameter, matter density parameter and density parameter for cosmological constant at present respectively. For a spatially flat Universe, $\Omega_{m0} + \Omega_{\Lambda 0} = 1$.
With such a background evolution, the equation (\ref{zero}) takes the form

\begin{equation}
a^2h^2(\frac{f_{,a}}{a}+f_{,a,a})=\frac{3}{2}\left(\frac{\Omega_{m0}}{a^3}f-\Omega_{\Lambda 0}\right)-af_{,a}\left(\frac{\Omega_{m0}}{2a^3}-\Omega_{\Lambda 0}\right).
\label{recons}
\end{equation} 

Here $f_{,a}$ represents the first derivative of $f$ with respect to the scale factor. One has to solve this equation to find the form for $f$ which can mimic the similar background evolution as in $\Lambda$CDM Universe. Equation (\ref{recons}) is a second order differential equation and needs two initial conditions, $f(initial)$ and $f_{,a}(initial)$ , to solve. We solve the equation from the era of decoupling ($ a \sim 10^{-3}$ ) till present day ($a=1$). We observe that the behavior does not vary with $f_{,a}(initial)$  but is very sensitive to $f(initial)$. In Figure 1, we show the different behaviors for $f(a)$  and $f'(a)$ with varying $f(initial)$ that give rise to the same background evolution as in $\Lambda$CDM model.

\begin{figure*}
\includegraphics[width=7cm]{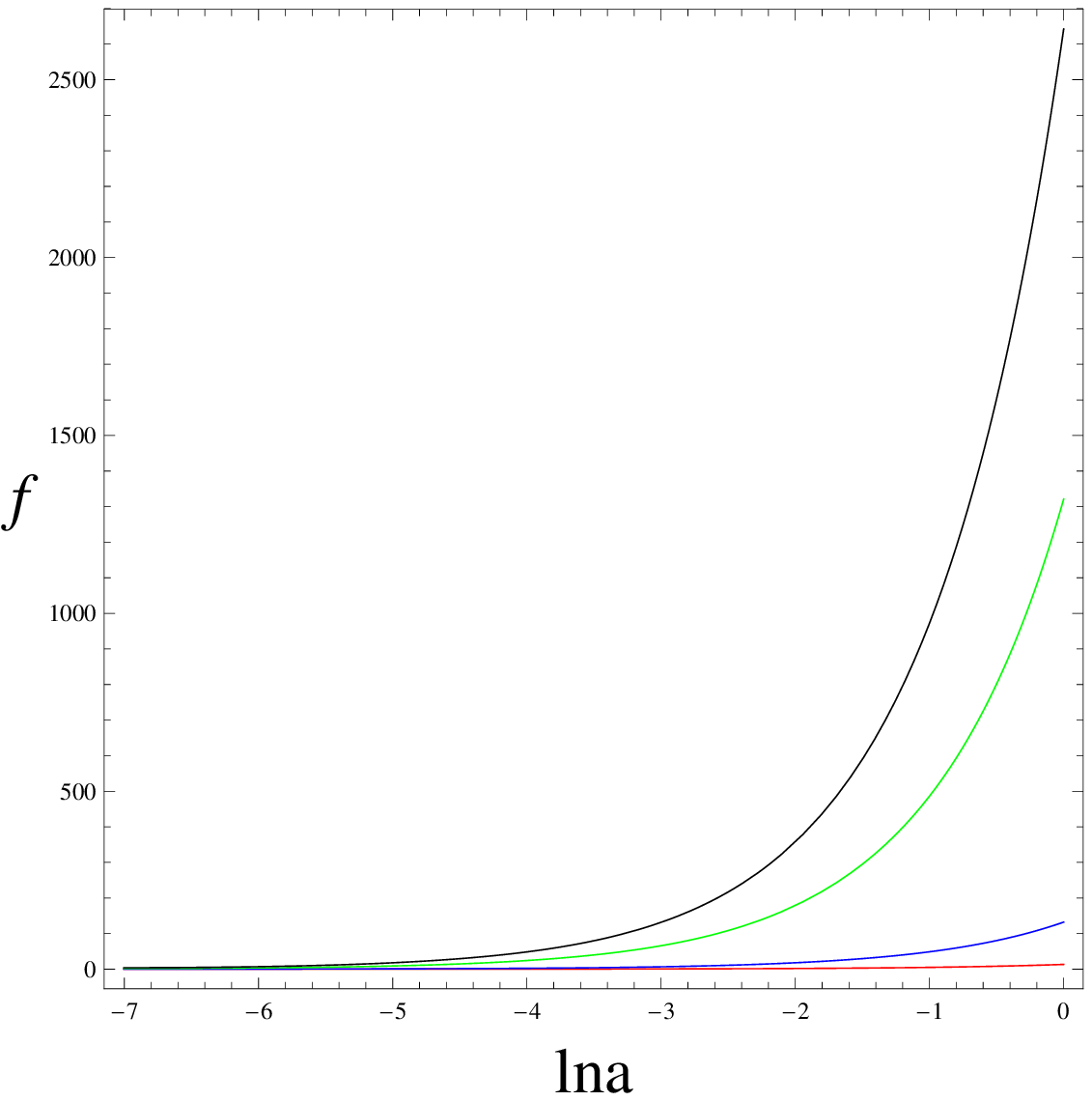}
\includegraphics[width=8cm]{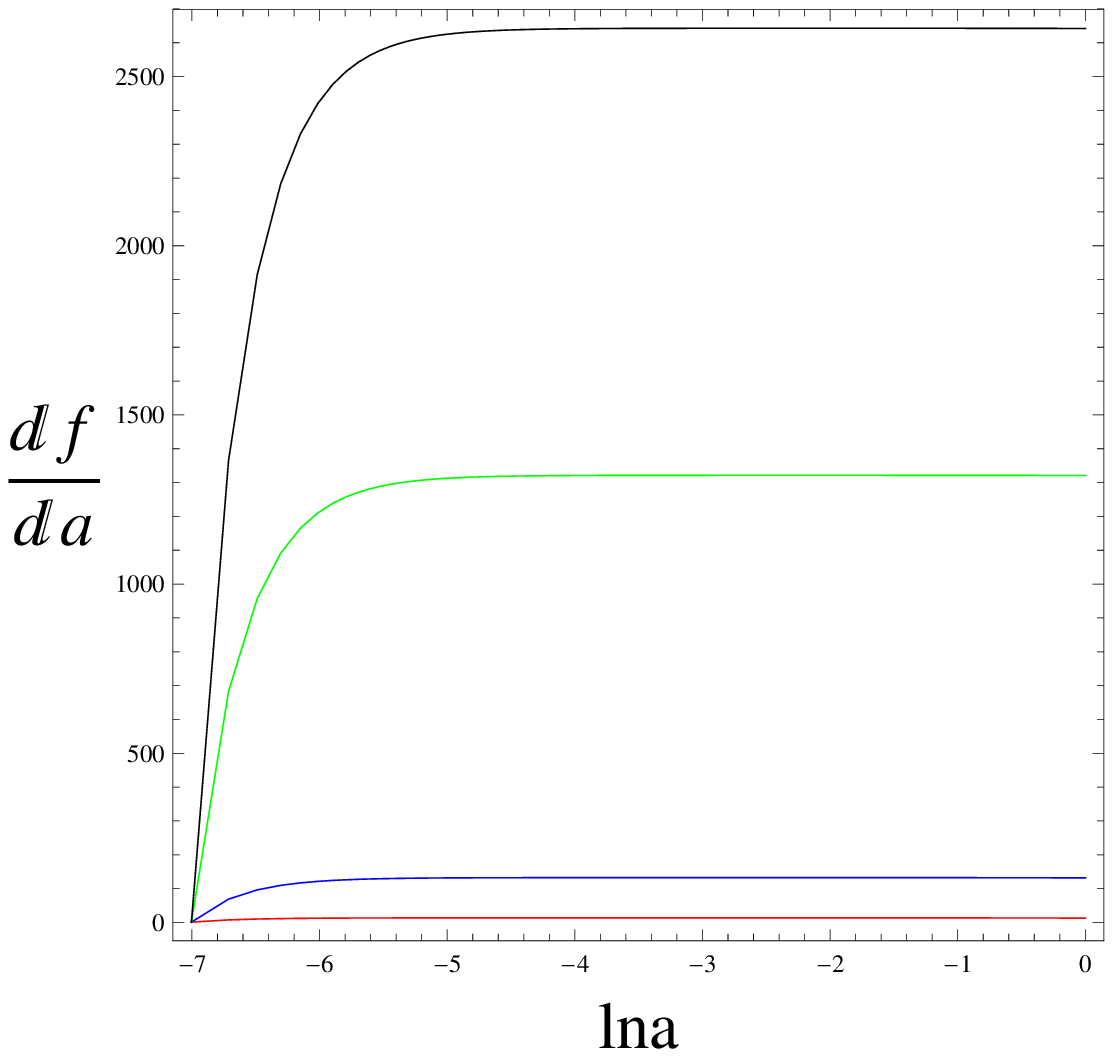}
\caption{Left figure shows behavior of $f$ with log of scale factor. Right figure corresponds to the behavior of $\frac{df}{da}$ with log of scale factor. In each figure different curves correspond to different values of parameter $f_{initial}$. From bottom to top $f_{initial}$ varies as $0.02, 0.2, 2.0, 4.0$ }
\end{figure*}

Given these forms for $f(R)$ that produce identical background evolution as in $\Lambda$CDM Universe, the next question is how different is the growth history of the Universe for these non-minimally coupled $f(R)$ gravity models from the concordance $\Lambda$CDM model? In the next section, we try to address this issue.

\section{Cosmological perturbations}

Perturbed FRW metric in longitudinal gauge takes the form:
\begin{equation}
ds^2=-(1+2\phi)dt^2+a^2(1-2\psi)\delta_{ij}dx^idx^j,
\label{perturb}
\end{equation}
where $\phi$ and $\psi$ are the two gravitational potentials. Using equations (\ref{eins}) and (\ref{perturb}), one can obtain the equations for growth in the linear regime:

\begin{eqnarray}
\frac{k^2}{a^2}\psi+3H(\dot{\psi}+H\phi)=-\frac{1}{2\chi}((3\dot{H}+3H^2-\frac{k^2}{a^2})\delta \chi  \nonumber \\-3H\dot{\delta \chi}+3H\dot{\chi}\phi+3\dot{\chi}(H\phi+\dot{\psi}) \nonumber \\+\kappa^2 (1+f)\delta\rho_m),
 \label{timecomp}
\end{eqnarray}

\begin{eqnarray}
\ddot{\delta \chi}+3H\dot{\delta \chi}+[\frac{1}{\chi_{,R}}(\frac{\chi}{3}+\kappa^2 \rho_m f_{,R})-\frac{R}{3}+\frac{k^2}{a^2}]\delta \chi=\nonumber \\   
\dot{\chi}(3H\phi+3\dot{\psi}+\dot{\phi})+(2\ddot{\chi}+3H\dot{\chi})\phi+\frac{\kappa^2}{3}(1+f)\delta\rho_m,
\label{tracecomp}
\end{eqnarray}
\begin{equation}
\psi-\phi = \frac{\delta \chi}{\chi} \label{diagonal}.
\end{equation}
Here $\delta\rho_{m}(t,\vec{x})= \rho_{m}(t,\vec{x}) -{ \rho}_{m}(t)$ and $\delta \chi(t,\vec{x}) = \chi(t,\vec{x})-{\chi}(t) $. Next we assume the velocity perturbation as
\begin{equation}
u^{\mu}  = u^{\mu (0)} + \delta u^{\mu}.
\end{equation}

\noindent
In the FRW Universe, $u^{\mu (0)} = (-1,0,0,0)$. Also with $u^{\mu}u_{\mu}= -1$, one can write $\delta u^{0} = - \delta u_{0} = \phi$. The spatial part of the $\delta u^{\mu}$ is the peculiar velocity. Writing the spatial part $\delta u^{i}$ as gradient of a scalar:
\begin{equation}
\delta u^{i} = \delta^{ij} v_{m,j},
\end{equation}

\noindent
one can now get the following equations on perturbing equation (2):
\begin{equation}
\dot{\delta}_m+{\bigtriangledown}^2 v_m-3\dot{\psi}=0,
  \label{eom1}
\end{equation}
\begin{equation}
\dot{v}_m+\left(2H+\frac{f_{,R}}{1+f}\dot{R} \right) v_m+\frac{1}{a^2}\left(\phi+\frac{f_{,R}}{1+f}\delta R\right)=0  \label{eom2}
\end{equation}
where $\delta_m = \frac{\delta \rho_m}{\rho_m}$. Using these two equations, one finally arrives at
\begin{eqnarray}
\ddot{\delta}_m + \left(2H+\frac{f_{,R}}{1+f}\dot{R}\right)\dot{\delta_m}+\frac{k^2}{a^2}\left(\phi+\frac{f_{,R}}{1+f}\delta R\right)=\nonumber \\3\ddot{\psi}+3\dot{\psi}\left(2H+\frac{f_{,R}}{1+f}\dot{R}\right). \label{eom3}
\end{eqnarray}

\subsection{Evolution of Matter Overdensities}

Equation (16) governs the cosmological evolution of the growth of matter over density in a non minimally coupled $f(R)$ gravity model which has identical background evolution as in the $\Lambda$CDM model. We are interested in the perturbations which are deep inside the Hubble radius for which $k  >> aH$.  We apply the following approximations to calculate the perturbations deep inside the Hubble radius \cite{tsuji} :
\begin{equation}
\{\frac{k^2}{a^2}|\phi|,\frac{k^2}{a^2}|\psi|,\frac{k^2}{a^2}|\delta \chi|\}>>\{H^2|\phi|,H^2|\psi|,H^2|\delta \chi|\}.
\end{equation}

\noindent
This actually implies $|\dot{Y}|\leq|HY|$, where $Y=\phi, \psi, \chi, \dot{\chi}, \delta \chi, \dot{\delta \chi}$ \\  Under this approximation, equations (9), (11) and (10) take the form

\begin{equation}
\frac{k^2}{a^2}\psi=\frac{1}{2\chi}\left(\frac{k^2}{a^2}\delta \chi- \kappa^2(1+f)\delta \rho_m\right)
\end{equation}
\begin{figure}[t]
\includegraphics[width=8cm]{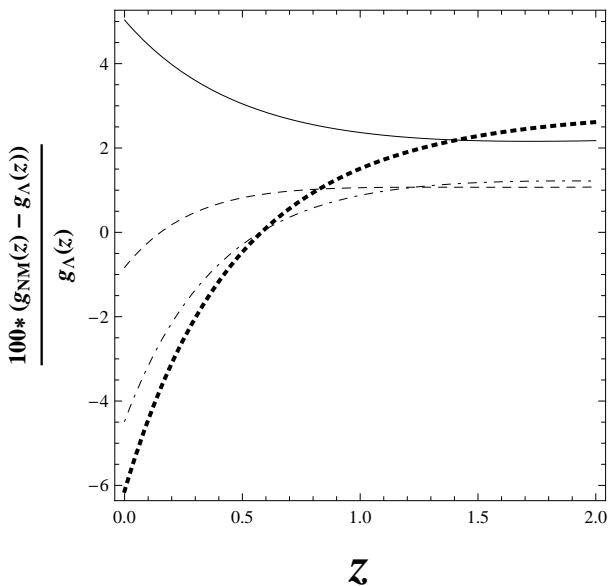}
\caption{Percentage deviation for the growth $g(z)$ from the corresponding $\Lambda$CDM value for different $f(initial)$. The solid, dashed, dot-dashed and dotted lines are for $f(initial) = 0.002, 0.006, 0.02, 0.2$ respectively. $\Omega_{\Lambda 0} = 0.7$.}
\end{figure}

\begin{equation}
\frac{k^2}{a^2}\phi=-\frac{1}{2\chi}\left(\frac{k^2}{a^2}\delta \chi+ \kappa^2(1+f)\delta \rho_m\right)  \label{nwtimecomp}
\end{equation}

\begin{equation}
\left[\frac{1}{\chi_{,R}}(\frac{\chi}{3}+\kappa^2 \rho_m f_{,R})-\frac{R}{3}+\frac{k^2}{a^2} \right]\delta \chi=\frac{\kappa^2}{3}(1+f)\delta \rho_m \label{nwtrace}.
\end{equation} 

Using equations (18), (19) and (20), and under the approximation (17), equation (16) finally becomes
\begin{equation}
\ddot{\delta}_m+\left(2H+\frac{f_{,x}}{1+f}\frac{\dot{R}}{R_0}\right)\dot{\delta}_m -4\pi G_{eff}\rho_m \delta_m=0 \label{final},
\end{equation}
where
\begin{equation}
\hspace{-0.5cm} \frac{G_{eff}}{G}=\frac{(1+f)\left[(\frac{\beta f_{,x}-1}{\beta f_{,x,x}})(1-\frac{2}{\alpha a^2}\frac{k^2}{H_0^2}\frac{f_{,x}}{1+f})-\frac{3}{2}\frac{f_{,x}}{f_{,x,x}}-x+\frac{4}{\alpha a^2}\frac{k^2}{H_0^2}\right]}{(\frac{\beta f_{,x}-1}{\beta f_{,x,x}})(1+\frac{3\Omega_{0m}}{\alpha a^3})f_{,x}+x(\beta f_{,x}-1)-\frac{3}{\alpha a^2}\frac{k^2}{a^2}(\beta f_{,x}-1)}
\end{equation}
Where $\beta = \frac{6 \Omega_{m0}}{\alpha a^3}$ and $\alpha = \frac{R_{0}}{H_{0}^2}$.  To solve the equation (21), we use the initial conditions  $\delta(a_{initial})=a_{initial}$ and $\delta ^{'}(a_{initial})=1$ where we choose $a_{initial} \sim 10^{-3}$, which is the decoupling era. This is to ensure that the growth evolution reproduces the matter-like behavior during the early time.
\begin{figure*}[t]
\includegraphics[width=8cm]{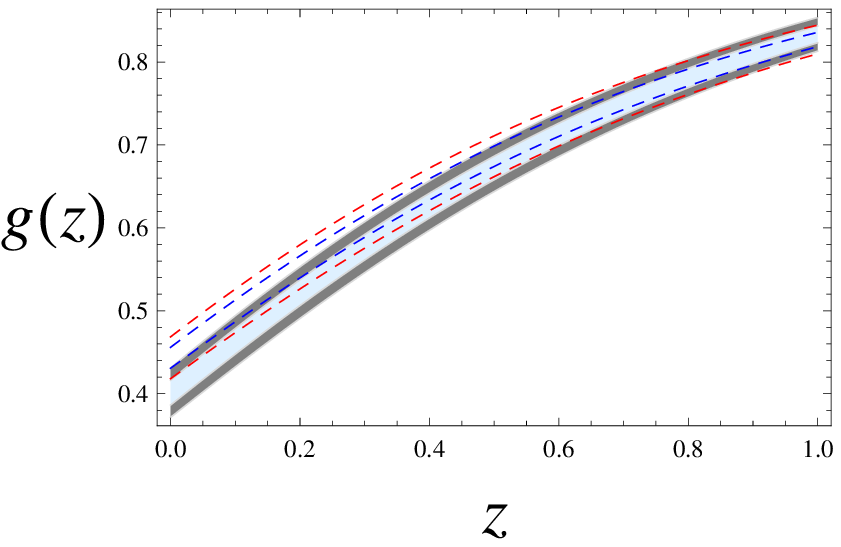}
\includegraphics[width=8cm]{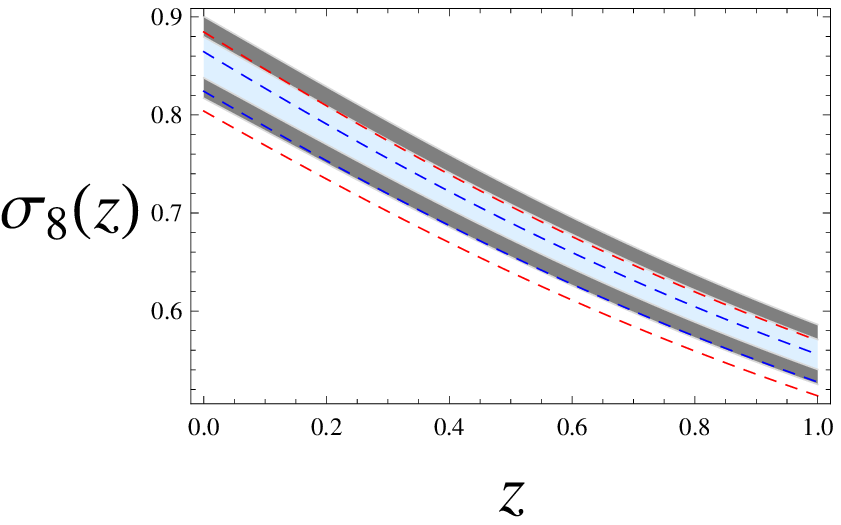}
\caption{Left figure corresponds to the growth rate of matter densities with respect to redshift. Right figure corresponds to the $\sigma_8$. The inner, outer shaded region corresponds to $1\sigma$ and $2\sigma$ deviation from the best fit value of parameters corresponding to non minimally coupled $f(R)$. The inner, outer dashed curves corresponds to $1\sigma$ and $2\sigma$ deviation from the best fit value of parameters corresponding to $\Lambda CDM$ }
\end{figure*}
\section{Observational Constraints}

Growth of matter overdensity  is an important independent probe for different cosmological models that explain the late time acceleration of the Universe. The growth rate for matter over density is defined as
\begin{equation}
g \equiv \frac{dln\delta_{m}}{dlna}.
\end{equation}

In Figure 2, we show the percentage deviation of the growth factor $g$ for the non-minimally coupled $f(R)$ gravity model from the corresponding $\Lambda$CDM model.   One can see that there can be four to six percent deviation one may expect for different values of $f(initial)$. The question  is whether such deviation can be probed by the current observational data.

Another important quantity related to the growth is the rms fluctuation of the linear density field at the scale  $8h^{-1}$Mpc, $\sigma_8(z)$. This defined as

\begin{equation}
\sigma_8 ^2(R,z)=\int_0 ^{\infty} \Delta^2(k,z) W(kR) \frac{dk}{k},
\end{equation}
 where $W(kR)$ is the window function defined as
  \begin{equation}
W(kR)=3\left(\frac{Sin(kR)}{(kR)^3}-\frac{Cos(kR)}{(kR)^2}\right).
\end{equation}

The quantity $\Delta^2(k,z)$ as defined in \cite{dragan}
\begin{equation}
\Delta^2(k,z)=A_s \frac{4}{25}\frac{1}{\Omega_m ^2}\left(\frac{k}{k_0}\right)^{n_s -1} \left(\frac{k}{H_0}\right)^4 D(z)^2 T(k)^2,
\end{equation}
Where $D(z)$ is the normalized density contrast defined as $\frac{\delta_m (z)}{\delta_m (z=0)}$, $A_s$ is the amplitude of the primordial curvature perturbation produced during inflation, $k_0$ = 0.05 $Mpc^{-1}$ is the pivot scale at which the primordial fluctuation is calculated.  $T(k)$ is the transfer function which incorporates the effects of evolution of perturbations through horizon crossing and matter/radiation transition. We use analytical form for  $T(k)$ as proposed in \cite{bbks}:
\begin{eqnarray}
T(k)=\frac{ln(1+2.34q)}{2.34q} \nonumber \\ \left[ 1+3.89q+(16.2q)^2+  (5.47q)^3+(6.71q)^4\right]^{-0.25}
\end{eqnarray}

Where $q \equiv \frac{1}{\Gamma h}\frac{k}{ Mpc^{-1}}$ 
and $ \Gamma = \Omega_{m0} h \hspace{0.1cm}exp^{- \Omega_{b0}(1+\sqrt{\frac{2h}{\Omega_{m0}}}) }$.

Galaxy surveys are directly sensitive to the combination $g(z) \sigma_8(z)$. This combination is almost a model independent estimator for the observed growth history of the Universe and that is why most of the surveys e.g. the 2dF, VVDS, SDSS, 6dF, BOSS, as well as the Wiggle-Z galaxy survey provide the measurement for this estimator. In a recent paper \cite{tabel}  compilation of the current measurements for $g(z) \sigma_8(z)$ has been given and we use those measurements for our purpose.

Using these measurements,  we calculate the posterior for our case, which is given by 
\begin{equation}
Posterior= P(\theta) {\cal L}(\theta)
\end{equation}
where $\theta$ corresponds to parameters in our theory. $P(\theta)$ is the prior probability distribution for different parameters. In our calculations, we fix the value of the spectral index $n_{s}=0.9616$ which is the best fit value obtained by the Planck \cite{planck}. This is because, posterior is not very sensitive to the value of $n_{s}$. We use  Gaussian prior for the parameters $\Omega_{\Lambda0}$  and $A_{s}$, with their valuse as $\Omega_{\Lambda0} = 0.686 \pm 0.020 $  and $ (10^9 A_{s}) = 2.23 \pm 0.16 $ from recent measurement by Planck \cite{planck}. For the model parameter $f (initial)$, we use a flat prior between $0.002$ to $4$. The likelihood function is defined as
\begin{equation} 
{\cal L}(\theta)= e^{-\frac{\chi(\theta)^2}{2}}.
\end{equation}
with
\begin{equation}
\chi^2= \Sigma \left( \frac{{g\sigma_{8}}_{obs}(z_i)-{g\sigma_{8}}_{th}(z_i,\Omega_{\Lambda},f_i,A_s)}{\sigma_{g\sigma_8}} \right)^2
\end{equation}

Covariance matrices for $f(R)$ with non minimal coupling and for $\Lambda CDM$ are constructed from the posterior. For non minimally coupled f(R) with parameters as $\Omega_{\Lambda}$, $f_i$ and $A_s$ it becomes  \\
 \[ C_{NM}= \left( \begin{array}{ccc}
 \Omega_{\Lambda 0} & f_{initial} & A_s \\
 0.000135109 & -0.000183621 & 0.00167062 \\
 -0.000183621 & 1.17092 & 7.28979*10^{-12} \\
 0.00167062 & 7.28979*10^{-12} & 0.0285568 \\
\end{array} \right) \]  
Covariance matrix for $\Lambda CDM$ with parameters as $\Omega_{\Lambda}$ and $A_s$ becomes
\[ C_{\Lambda CDM}= \left( \begin{array}{cc}
 \Omega_{\Lambda 0} & A_{s} \\
 0.000134686 & 0.0016725 \\
 0.0016725 & 0.0285589 \\
\end{array} \right) \] 

Using these covariance matrices of non minimally coupled $f(R)$ as well as for $\Lambda$CDM, we have reconstructed evolution of the growth  factor $g(z)$ and $\sigma_8(z)$ at $1\sigma$ and $2\sigma$ confidence level. This has been shown in Figure 3.  One can see clearly that it is possible to distinguish the $\Lambda$CDM and the non-minimal $f(R)$ model through the $g(z)$ evolution both at $1\sigma$ and $2\sigma$ confidence level, whereas for the  $\sigma_8(z)$ evolution, one can distinguish them at $2\sigma$ confidence level.  We should stress the fact that two models have identical background  evolution and can not be distinguished by any observation related to background cosmology.

\section{Conclusion}

Explaining late time acceleration is one of the most significant challenges for cosmologists today. Modifying the Einstein gravity at large cosmological scales like $f(R)$ gravity models, is one interesting approach to explain such late time acceleration. Coupling curvature with matter in $f(R)$ gravity has been recently proposed which has the advantage, that 
the singularity that occurs in the standard $f(R)$ gravity models may not occur in non minimally coupled $f(R)$. The background cosmology in such non-minimally coupled $f(R)$ gravity has been studied. Recently its implication on inhomogeneous Universe has also been explored in \cite{berto}.  

Our goal is to extend further the work done in \cite{berto} to investigate the possibility to distinguish  such a non minimally coupled $f(R)$ gravity model from the concordance $\Lambda$CDM  model using the current measurements by different galaxy surveys. For this we do not assume any particular form for the function $f(R)$.  We take a different approach in this investigation.  The growth of matter overdensity in any modified gravity model depends both on the background expansion as well as the extra effects that arise solely due to the modification in the gravity action ( in this case it is the non minimal coupling between the curvature and matter). To study the effect of this non minimal coupling on the growth history and the subsequent deviation from the $\Lambda$CDM behavior, we fix the background evolution same as the that of the $\Lambda$CDM model thereby eliminating the effect due to the background evolution. Now departure from the $\Lambda$CDM evolution is solely due to the modified equation for the growth evolution in the non-minimally coupled case. This is one crucial difference in approach from the investigation done in \cite{berto} where a specific form for $f(R)$ was assumed.

Subsequently we calculate the allowed behaviors for $f(R)$ modification, that can give rise to the same background cosmology as in $\Lambda$CDM Universe. Reconstructing this form, we subsequently study the growth history and the $\sigma_{8}$ normalization for such $f(R)$ forms and compare them with $\Lambda$CDM Universe. We use the measurements for $g(z)\sigma_{8}(z)$ ( which is mostly model independent) from different galaxy surveys to reconstruct the form for growth  factor $g(z)$ and $\sigma_{8}(z)$ for both the non minimally coupled $f(R)$ model as well as for $\Lambda$CDM Universe. The results show that it is possible to distinguish these two models using the growth of matter over density, even if they have same background evolutions. This result is the important extension of the work done in \cite{berto} to observationally distinguish the non minimally coupled $f(R)$ gravity model from the concordance $\Lambda$CDM model.

We should mention that we have considered the simple form for the action as given in equation (1). One can consider more general form like ${\cal L} = f_{1}(R) + (1+f_{2}(R)){\cal L}_{m}$. In that case, if one considers the same background evolution as in $\Lambda$CDM, there will be effects on the growth history from the modified part of the pure gravity action ($f_{1}(R)$) as well as the effect due to the non minimal coupling between the matter and the curvature.  One can then address two questions. Firstly whether such model can be distinguished from the concordance $\Lambda$CDM using the current observational data. And also given a specific form for the pure gravity lagrangian $f_{1}(R)$, whether the non minimal $f(R)$  model can be distinguished from the standard minimal $f(R)$ model using the observational data. We shall address these issues in future.

\section{Acknowledgement}
The authors thank T.R. Seshadri for useful discussions and comments. ST acknowledges the computational facilities provided by the CTP, JMI, New Delhi, India.  ST thank C.S.I.R, Govt. of India for financial support through Senior Research Fellowship. AAS. acknowledges the partial financial support provided by C.S.I.R. Govt. of India through the research grant (Grant No:03(1187)/11/EMR-II).

\end{document}